\documentclass[runningheads]{llncs}

\usepackage{graphicx}

\usepackage{amsmath}
\usepackage{mathtools}
\usepackage{url}

\usepackage{multirow}
\usepackage{booktabs}
\usepackage{subcaption}
\usepackage{pifont}
\newcommand{\cmark}{\ding{51}}
\newcommand{\xmark}{\ding{55}}
\usepackage{hyperref}

\begin{document}

\title{Recognition of food-texture attributes using an in-ear microphone%
  \thanks{Part of this work has been presented in the first author's
    Ph.D. thesis \cite{papapanagiotou2019phd}.%
    \hfill\break%
    The work leading to these results has received funding from (a) the European
    Community's ICT Programme under Grant Agreement No. 610746, 01/10/2013 -
    30/09/2016 \url{https://splendid-program.eu/}, and (b) the European
    Community's Health, demographic change and well-being Programme under Grant
    Agreement No. 727688, 01/12/2016 - 30/11/2020 \url{https://bigoprogram.eu}.%
  }%
}


\author{Vasileios Papapanagiotou\inst{1}\orcidID{0000-0001-6834-5548}%
  \and Christos Diou\inst{1,2}\orcidID{0000-0002-2461-1928}%
  \and Janet van den Boer\inst{3}\orcidID{0000-0001-7526-376X}%
  \and Monica Mars\inst{4}\orcidID{0000-0003-2010-1039}%
  \and Anastasios Delopoulos\inst{1}\orcidID{0000-0001-8220-8486}}

\authorrunning{V. Papapanagiotou et al.}

\institute{%
  Multimedia Understanding Group, Dpt. Electrical and Computer Engineering,
  Faculty of Engineering, Aristotle University of Thessaloniki, Greece\\
  \email{vassilis@mug.ee.auth.gr, adelo@eng.auth.gr}%
  \and Department of Informatics and Telematics,\\
  Harokopio University of Athens\\
  \email{cdiou@hua.gr}%
  \and Dpt. of Biomedical Signals and Systems, Faculty of Electrical
  Engineering, Mathematics and Computer Science, University of Twente, The
  Netherlands\\
  \email{j.h.w.vandenboer@utwente.nl}%
  \and Division of Human Nutrition and Health,\\
  Wageningen University\\
  \email{monica.mars@wur.nl}}

\maketitle              

\begin{abstract}
  Food texture is a complex property; various sensory attributes such as
  perceived crispiness and wetness have been identified as ways to quantify
  it. Objective and automatic recognition of these attributes has applications
  in multiple fields, including health sciences and food engineering. In this
  work we use an in-ear microphone, commonly used for chewing detection, and
  propose algorithms for recognizing three food-texture attributes, specifically
  crispiness, wetness (moisture), and chewiness. We use binary SVMs, one for
  each attribute, and propose two algorithms: one that recognizes each texture
  attribute at the chew level and one at the chewing-bout level. We evaluate the
  proposed algorithms using leave-one-subject-out cross-validation on a dataset
  with 9 subjects. We also evaluate them using leave-one-food-type-out
  cross-validation, in order to examine the generalization of our approach to
  new, unknown food types. Our approach performs very well in recognizing
  crispiness (0.95 weighted accuracy on new subjects and 0.93 on new food
  types) and demonstrates promising results for objective and automatic
  recognition of wetness and chewiness.
\end{abstract}

\section{Introduction}
\label{sec:introduction}


Chewing is one of the main ways of how we perceive food texture
\cite{pereira2016}. While there is a huge variety of products with different
food textures (e.g. crispy products like potato chips, hard and moist products
like apples and cucumbers), some textures are generally perceived as more
pleasant and desirable than others \cite{cox1998}. It is clear from several
studies that food texture and structure are becoming more important in
understanding eating behavior, especially in food intake regulation and weight
management
\cite{lasschuijt2017comparison,zijlstra2009effect,aguayo2019oral,stribitcaia2020}. Taking
food structure into account in dietary advice can may support the prevention and
dietary treatment of overweight and obesity
\cite{campbell2017designing,boer2017availability,forde2018from}. However, the
effects still need to be supported by longer-term studies outside the
laboratory. The current, existing dietary assessment methods, such as diaries
and recalls, rely on memory and do not provided detailed information about the
texture of foods \cite{brouwer2020dietary}.


Currently, there is a strong effort in creating automated tools for dietary
monitoring, in the context of understand and preventing obesity and eating
disorders \cite{sazonov2010automatic,papapanagiotou2017novel}. One of the first
approaches has been to automatically detect chewing based on the audio captured
by an in-ear microphone \cite{amft2005analysis}. Such audio signals have also
been used to extract information such as the food type
\cite{amft2010wearable,passler2012food}. There are also alternative types of
approaches for identifying food types; for example, leveraging photos that
people take with their mobile phones can be used to detect food-relevant photos
and then subsequently to perform image segmentation and identify the food type
\cite{lu2018multi,dehais2016food,christodoulidis2015food}. Alternatively, user
input can be requested when an automatic eating detection system detects eating
activity \cite{boer2018splendid}.

Identifying food-content--relevant information such as food type from audio
signals is commonly formulated as a multi-class problem where each class is a
different food type from a list of pre-determined food types
\cite{amft2005analysis,amft2009bite}. The selection of food types is usually
aimed at creating a diverse set with different textures, such as crispy and
non-crispy food types; crispiness, however, is only one of the attributes of
texture. In the early work of \cite{munoz1986development}, a set of texture
reference scales were introduced that include multiple attributes. Table
\ref{tab:food_attributes_munoz} presents three groups of attributes: attributes
related to surface attributes and springiness, attributes assessed during
mastication (chewing), and those assessed during manual manipulation. Out of the
three groups, the attributes assessed during mastication are the ones of
interest to this work. The more recent work of \cite{szczesniak2002texture}
presents a more extensive and complete review of the state of knowledge for food
texture. According to that work, it is commonly accepted that ``texture is the
sensory and functional manifestation of the structural, mechanical and surface
properties of foods detected through the senses of vision, hearing, touch and
kinesthetics''. As a result, no single modality sensor (such as a microphone)
can completely identify texture.

\begin{table}
  \centering
  \caption{Food-texture attributes as presented and organized in
    \cite{munoz1986development}, and their correspondance with the
    food-attributes used in this work.}
  \label{tab:food_attributes_munoz}
  \begin{tabular}{lccc}
    \toprule
    \textbf{Attributes} & \textbf{Crisp} & \textbf{Wet} & \textbf{Chewy}\\
    \midrule
    \multicolumn{4}{l}{\emph{Attributes related to surface attributes and springiness}}\\
    wetness & \xmark & \cmark & \xmark \\
    adhesiveness to lips & \xmark & \xmark & \xmark \\
    roughness & \cmark & \xmark & \xmark \\
    self-adhesiveness & \xmark & \xmark & \cmark \\
    springiness & \cmark & \xmark & \xmark \\
    \midrule
    \multicolumn{4}{l}{\emph{Attributes assessed during mastication}}\\
    cohesiveness of mass & \cmark & \xmark & \xmark \\
    moisture absorption & \xmark & \cmark & \xmark \\
    adhesiveness to teeth & \xmark & \xmark & \cmark \\
    \midrule
    \multicolumn{4}{l}{\emph{Attributes assessed during manual manipulation}}\\
    manual adhesiveness & \xmark & \xmark & \xmark \\
    \bottomrule
  \end{tabular}
\end{table}

It is worth mentioning that there are additional, non-medical fields where
research in understanding human eating behavior is also useful. More
specifically, in the field of food design and engineering, understanding how
people perceive certain attributes of food (such as crispiness) has been found
to correlate with freshness (in particular for the case of apples
\cite{daillant1996relationships,peneau2007relating}) which in turn is the most
important factor in consumer's choices
\cite{harker2003case,peneau2006importance}. Thus, providing tools that can
objectively measure such food attributes cannot only help to assess eating
behavior and food intake, but can also help food technologists to design food
with more desirable and pleasant characteristics.

In this work we use an in-ear microphone that is part of the wearable, prototype
sensor developed in the context of the SPLENDID project and focus on the audio
captured during chewing in order to recognize three attributes of food texture
from a variety of food types. The three attributes are crispiness, wetness, and
chewiness. Each one corresponds to each of the attributes ``assessed during
mastication'' according to \cite{munoz1986development} (table
\ref{tab:food_attributes_munoz}). The attributes ``related to surface attributes
and springiness'' can also be loosely mapped to these three attributes. The
attributes that are ``assessed during manual manipulation'' are not of interest
to this work (but are only included in the table for completeness).

We propose an algorithm for recognizing each of the three food-texture
attributes of this work (i.e. crispiness, wetness, and chewiness) given a single
chew based on three individual binary SVMs (one for each attribute). We also
propose a modified version of the algorithm that operates on entire chewing
bouts. We evaluate the generalization both in new subjects and in new food
types.

\section{Attribute recognition algorithms}

The algorithms presented in this work require audio of at least $8$ kHz sampling
rate. The original audio recordings of the dataset used in this work (see
Section \ref{sec:food_datasets}) have been sampled at $48$ kHz. We have
experimented with down-sampled versions of the signal, in particular $2$, $4$,
$8$, $16$, $32$, as well as the original $48$ kHz. We have observed that
down-sampling as low as $8$ kHz does not cause any noticeable drop in
recognition effectiveness (by repeating the experiments presented in this paper
for all different sampling frequencies), however, down-sampling lower than $8$
kHz does. In all the following, we use the $8$ kHz down-sampled versions of the
audio signals.

We also apply a high-pass Butterworth filter to the down-sampled signal to
remove low frequency components. The filter is of $9$-th order with a cut-off
frequency at approximately $20$ Hz. We propose two algorithms: the first
recognizes attributes for each chew individually, while the second one for
entire chewing bouts.

\subsection{Chew-level algorithm}
\label{sec:algorithm1}

A feature vector is first extracted from each chew; note that start and stop
time-moment annotations for chews have been made available by manual extraction
based on acoustic and visual observation of the captured audio signals (in a
fully automated application, a chew-detection algorithm such as
\cite{papapanagiotou2017novel} can be used to obtain them). The extracted
features consist of signal energy in $11$ log-scale frequency bands based on
time-varying spectrum (TVS) estimation, fractal dimension (FD), condition number
(CN) of the auto-correlation matrix, and higher order statistics (i.e. third and
fourth-order moments). These features have been used in
\cite{papapanagiotou2017novel} for chewing detection and are thus a good
starting point for food-texture--attribute recognition. Since the audio sampling
frequency is only $2$ kHz in \cite{papapanagiotou2017novel} (compared to $8$ kHz
of this work) we have added two more bands in the TVS, in particular $2$ to $4$
and $4$ to $8$ kHz. It is also worth noting that each chew has different
duration (average of $0.56$ s and standard deviation of $0.15$ s in the data-set
used in this work) in contrast to windows of fixed length often used in signal
processing. The features we have selected, however, are invariant to the length
of the audio segment used to extract them.


Before the classification stage each feature is standardized by subtracting the
mean and then dividing by the standard deviation (the mean and standard
deviation of each feature are estimated over the available training set for each
case). The multi-label classifier we use is an array of three binary SVMs; each
SVM is related to one of the three food-texture attributes: crispiness, wetness,
and chewiness. We use a radial-basis function (RBF) kernel. Parameters $C$ and
$\gamma$ are selected automatically using $5$-fold cross-validation on the
training set. The optimal values are chosen using Bayesian optimization
\cite{gelbart2014bayesian}; care to escape local minima of the objective
function is also taken using a threshold for the standard deviation of the
posterior objective function \cite{bull2011convergence,snoek2012practical}.

As a result, each chew can be classified individually for each of the three
attributes. As the food type does not change within a bout, we can obtain one
decision per bout based on the chews that belong to it. One way to do so is to
use majority voting across the chews of a bout (for each attribute). Another way
is to consider only the first $n$ chews of a bout, since processing of the food
in the mouth transforms it into a wet bolus, thus altering the attributes of the
food in its unprocessed state. All these evaluation methods are presented in
Section \ref{sec:food_results}.

\subsection{Chewing-bout--level algorithm}

Chewing-bout--level detection shares the same pre-processing steps with
chew-level detection: the audio signal is down-sampled and the same high-pass is
applied to remove low-frequency components.

\begin{figure}
  \centering
  \includegraphics[scale=.8]{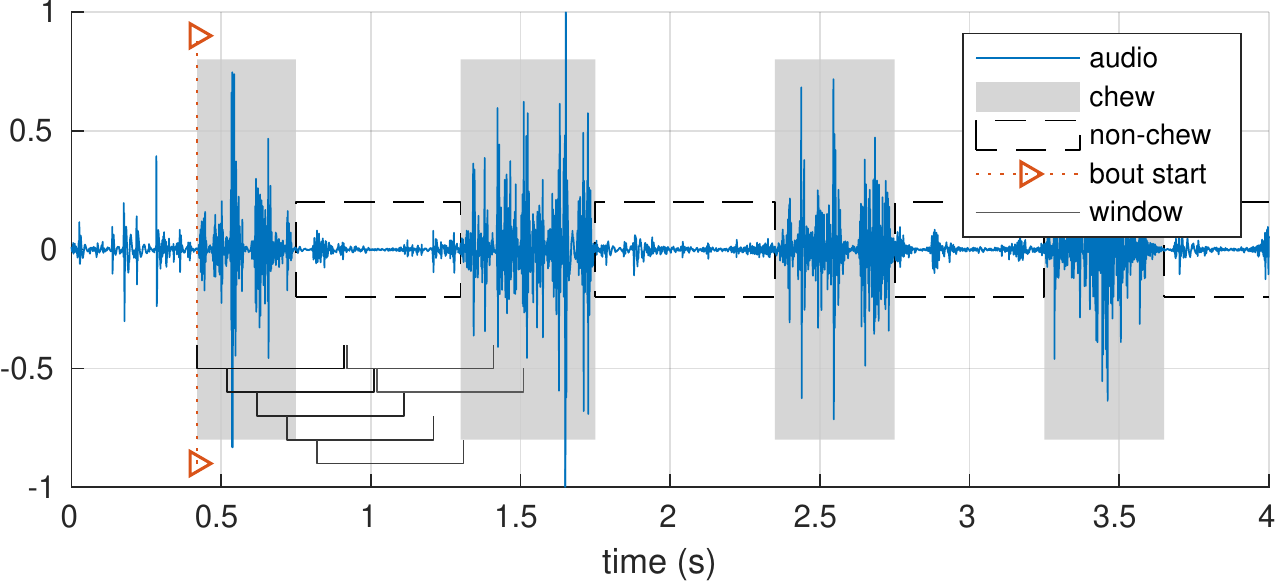}
  \caption{An example of a chewing bout. The first four chews are marked by the
    gray rectangles. When entire chewing bouts are used, the audio between two
    successive chews is also used.}
  \label{fig:audio_example}
\end{figure}

Bout segments are then obtained based on the chews that belong to each bout. A
bout audio segment starts at the start time of its first chew and stops at the
stop time of its last chew. The average bout duration in the data-set used in
this work is $15.22$ s with a standard deviation of $10.7$ s. Since bouts are
significantly longer than chews and also contain non-chewing sounds between each
chew (see Figure \ref{fig:audio_example}) we do not directly extract the
features over the entire bout duration. Instead, we obtain overlapping windows
of $0.5$ s length and $0.1$ s step from each bout and extract the features from
each window separately. Thus, we obtain one list of feature vectors for each
bout; each list contains, in general, a different amount of vectors.

We then use a bag-of-words (BoW) approach to obtain a single feature vector of
fixed length for each bout. In particular, given a set of bouts, we obtain all
of the feature vectors from each bout and use $k$-means to select a set of $k$
centroid vectors. Once we obtain the $k$ centroid vectors we can transform any
new bout into a feature vector of fixed length (equal to $k$) by computing the
normalized histogram of the bout's feature vectors against the set of $k$
centroid vectors. Each feature vector is assigned to one of the $k$ centroid
(i.e. hard assignment).

Using the BoW approach offers many advantages. It allows to use the same
features of the previous, chew-level algorithm (Section \ref{sec:algorithm1}) on
short windows (with similar duration to that of chews). In addition it allows to
extract feature vectors of fixed length from audio segments with (highly)
varying length. Finally, it also allows to handle the non-chewing sounds that
occur between successive chews within each bout: window-based feature vectors
that correspond to such in-between audio segments will likely be similar and
will be clustered together; corresponding cluster centers are equally present in
signals of different food types and the {SVM} models are expected to learn to
ignore them.

The BoW features are then standardized similarly to Section
\ref{sec:algorithm1}. Classification is performed in exactly the same way as in
chew-level recognition: we use an array of three binary {SVM} classifiers with
{RBF} kernel and hyper-parameter selection using Bayesian optimization.

\section{Data-set and evaluation metrics}
\label{sec:food_datasets}

The data-set used to evaluate the proposed algorithms has been collected at
Wageningen University, Netherlands, in the context of the EU-funded SPLENDID
project\footnote{\url{https://splendid-program.eu/}} and is the same data-set as
the one we use in \cite{papapanagiotou2015fractal}. The recording apparatus is
an in-ear microphone (FG-23329-D65 model manufactured by Knowles) connected via
wire to a computer audio interface. The sensor housing and recording has been
done by CSEM S.A. \cite{papapanagiotou2017novel}. In this work, the first
version of the SPLENDID sensor is used; more details about this version and the
future versions of the sensor can be found in \cite{boer2018splendid}. In total,
$21$ subjects were enrolled for the data collection trials, however, signals
from only $9$ could be used in this work due to problems with data acquisition
(such as incorrect sensor placement or corrupted audio due to hardware/software
malfunction). Each subject consumed a variety of food types (complete list can
be found in \cite{boer2018splendid}).

We have selected $9$ different food types that we can clearly annotate their
attributes. Not all $9$ subjects have consumed all $9$ food types. Table
\ref{tab:food_datasets_attributes} lists these food types along with their
attribute values we have assigned them. This data-set of $9$ participants and
$9$ food types is the one we use to evaluate this work.

\begin{table}
  \centering
  \caption{List of food types and their attributes for the evaluation data-set.}
  \label{tab:food_datasets_attributes}
  
  \begin{tabular}{lccc}
    \toprule
    & \textbf{Crispy} & \textbf{Wet} & \textbf{Chewy}\\
    \midrule
    apple          & \cmark & \cmark & \xmark\\
    banana         & \xmark & \cmark & \xmark\\
    bread          & \xmark & \xmark & \xmark\\
    candy bar      & \xmark & \xmark & \cmark\\
    cookie         & \cmark & \xmark & \xmark\\
    lettuce        & \cmark & \cmark & \xmark\\
    potato chips   & \cmark & \xmark & \xmark\\
    strawberry     & \xmark & \cmark & \xmark\\
    toffee         & \xmark & \xmark & \cmark\\
    \bottomrule
  \end{tabular}
\end{table}

For this evaluation data-set, we have manually created ground truth on chew and
chewing bout levels (with start and stop time-moments) based on the available
experimental logs as well as audio and visual inspection of the captured
signals. It contains $4,989$ chews with a total duration of $46.31$ min which
belong to $238$ chewing bouts; total duration of bouts is almost $1$ h and is
greater than total duration of chews because bouts also contain the audio
segments between each successive pair of chews. Tables
\ref{tab:food_datasets_chews_stats} and \ref{tab:food_datasets_bouts_stats} show
duration statistics for the chews and chewing bouts.

\begin{table}
  \centering
  \caption{Statistics of chews duration across the food types for the evaluation
    data-set.}
  \label{tab:food_datasets_chews_stats}
  
  \begin{tabular}{lrrrr}
    \toprule
    & \textbf{Instances} & \textbf{Total (min)}
    & \textbf{Average (s)} & \textbf{Std (s)}\\
    \midrule
    apple          & $1,091$ &  $9.74$ & $0.54$ & $0.13$\\    
    banana         &   $294$ &  $2.67$ & $0.54$ & $0.16$\\
    bread          & $1,260$ & $11.72$ & $0.56$ & $0.13$\\
    candy bar      &   $415$ &  $4.26$ & $0.62$ & $0.17$\\
    cookie         &   $359$ &  $3.07$ & $0.51$ & $0.14$\\
    lettuce        &   $492$ &  $4.06$ & $0.50$ & $0.11$\\    
    potato chips   &   $386$ &  $3.45$ & $0.54$ & $0.12$\\
    strawberry     &   $236$ &  $2.04$ & $0.52$ & $0.13$\\
    toffee         &   $456$ &  $5.30$ & $0.70$ & $0.17$\\
    \midrule
    total          & $4,989$ & $46.31$ & $0.56$ & $0.15$\\
    \bottomrule
  \end{tabular}
\end{table}

\begin{table}
  \centering
  \caption{Statistics of bouts duration across the food types for the evaluation
    data-set.}
  \label{tab:food_datasets_bouts_stats}
  
  \begin{tabular}{lrrrr}
    \toprule
    & \textbf{Instances} & \textbf{Total (min)}
    & \textbf{Average (s)} & \textbf{Std (s)}\\
    \midrule
    apple              & $48$       & $12.12$       & $15.15$       & $5.23$\\
    banana             & $25$        & $3.57$        & $8.58$       & $2.91$\\
    bread              & $37$       & $15.03$       & $24.37$      & $15.69$\\ 
    candy bar          & $22$        & $5.51$       & $15.02$       & $7.97$\\
    cookie             & $16$        & $4.03$       & $15.10$       & $5.31$\\
    lettuce            & $28$        & $5.31$       & $11.38$       & $5.39$\\
    potato chips       & $29$        & $4.40$        & $9.11$       & $3.12$\\
    strawberry         & $17$        & $2.63$        & $9.30$       & $5.82$\\ 
    toffee             & $16$        & $6.23$       & $23.35$      & $11.37$\\
    \midrule
    total             & $238$       & $58.83$       & $14.83$      &  $9.84$\\  
    \bottomrule
  \end{tabular}
\end{table}



We evaluate each food attribute, namely crispiness, wetness, and chewiness, as
binary classification problems. We regard crispy, wet, and chewy as the positive
classes, and non-crispy, dry, and non-chewy as the negative ones respectively.
To account for class imbalance, which is particularly large for chewiness, we
calculate weighted accuracy as
\begin{equation}
  \label{eq:wacc}
  \frac{w \cdot TP + TN}{w \cdot (TP + FN) + TN + FP}
\end{equation}
where $w = (TN + FP) / (TP + FN)$ is the ratio of priors.

\section{Evaluation \& Results}
\label{sec:food_results}

The chew-level algorithm can also be modified to operate on bout-level by taking
into account each bout's chews. We explore how the bout-level decision is
affected by the number of chews that are taken into account. We train models
both in the typical leave-one-subject-out (LOSO) fashion as well as in
leave-one-food-type-out (LOFTO) fashion. In both types of experiments we use the
entire available training data (from the non--left-out subjects or food types) to
both obtain the BoW centroid vectors and train the SVM classifiers.

Table \ref{tab:food_results_chews_loso} presents per-chew classification results
for each of the three attributes; the BoW centroids, {SVM} models, and SVM
hyper parameters have been trained in {LOSO} fashion. Crispiness is the
attribute that the algorithm identifies more effectively. The majority voting
approach consistently improves results by $2$ to $5\%$.

\begin{figure}
  \centering

  \begin{subfigure}{.48\linewidth}
    \centering
    \includegraphics[scale=.8]{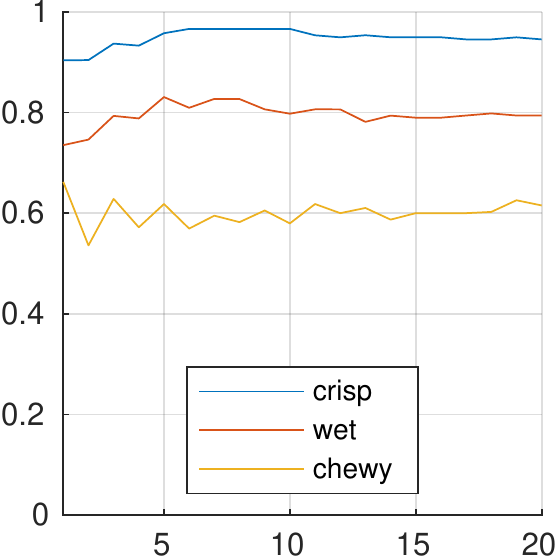}
    \caption{LOSO}
  \end{subfigure}
  \begin{subfigure}{.48\linewidth}
    \centering
    \includegraphics[scale=.8]{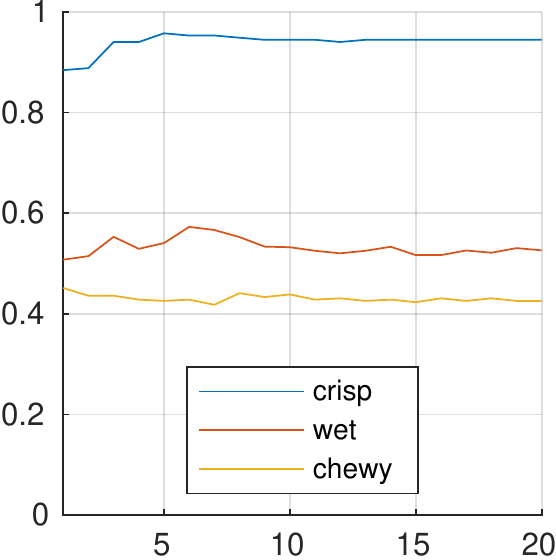}
    \caption{LOFTO}
  \end{subfigure}

  \caption{Weighted accuracy for each attribute for the LOSO and LOFTO
    experiments.}
  \label{fig:chews_majority}
\end{figure}

While it makes sense to assume that food type as such does not change during a
bout, the attributes of the food within the mouth do as food is grinded and
lubricated during oral processing/chewing.
Given that we are interested in identifying the attributes that the food has in
its unprocessed form, we can consider only the first few chews of each bout in
the majority voting stage, during which the food type's original attributes are
still retained to some degree. Figure \ref{fig:chews_majority} shows the
recognition effectiveness for each attribute when considering only the first $n$
chews of each bout for $n=1$ to $20$. Recognition of crispiness exhibits high
effectiveness, however, the highest (weighted) accuracy is obtained by
considering the first $6$ to $10$ chews. Almost the same range of $5$ to $10$
chews seems to be the best choice for recognizing wetness as well. Chewiness
results are somewhat different since considering only the first few chews seems
to yield erratic effectiveness. The situation improves as more chews are taken
into account.

\begin{table}
  \centering
  \caption{{LOSO} results for chew-level recognition.}
  \label{tab:food_results_chews_loso}
  \begin{tabular}{lccc}
    \toprule
    & \textbf{Prior} & \textbf{Weighted accuracy} & \textbf{$w$}\\
    \midrule
    \emph{Chew level}\\
    crispy (avg) & $0.4707$  & $0.9068$    & $1.1968$\\
    crispy (sum) & $0.4666$  & $0.9017$    & $1.1430$\\
    wet (avg)    & $0.4280$  & $0.7516$    & $1.4588$\\
    wet (sum)    & $0.4235$  & $0.7503$    & $1.3611$\\
    chewy (avg)  & $0.1741$  & $0.5994$    & $4.9418$\\
    chewy (sum)  & $0.1746$  & $0.6212$    & $4.7279$\\
    \midrule
    \multicolumn{3}{l}{\emph{Majority voting per bout}}\\
    crispy (avg) & $0.4943$  & $0.9519$    & $1.1141$\\    
    crispy (sum) & $0.5063$  & $0.9496$    & $0.9752$\\
    wet (avg)    & $0.4850$  & $0.7978$    & $1.1018$\\
    wet (sum)    & $0.4937$  & $0.7900$    & $1.0254$\\
    chewy (avg)  & $0.1666$  & $0.6296$    & $5.9804$\\
    chewy (sum)  & $0.1632$  & $0.6154$    & $5.1282$\\    
    \bottomrule
  \end{tabular}
\end{table}

Table \ref{tab:food_results_chews_lofo} presents results for the LOFTO
experiments. Crispiness is the most easily recognizable attribute again,
however, wetness does not generalize well across different food types. Majority
voting improves results for crispiness and chewiness only. Looking at figure
\ref{fig:chews_majority}, the highest effectiveness for wetness is achieved by
considering only a few chews ($6$ to $7$ chews).

\begin{table}
  \centering
  \caption{{LOFTO} results for chew-level recognition.}
  \label{tab:food_results_chews_lofo}
  \begin{tabular}{lcccc}
    \toprule
    & \textbf{Prior} & \textbf{Weighted accuracy} & \textbf{$w$}\\
    \midrule
    \emph{Chew-level}\\
    crispy (sum) & $0.4666$  & $0.8987$    & $1.1430$\\
    wet (sum)    & $0.4235$  & $0.5481$    & $1.3611$\\
    chewy (sum)  & $0.1746$  & $0.3957$    & $4.7279$\\
    \midrule
    \multicolumn{2}{l}{\emph{Majority voting per bout}}\\
    crispy (sum) & $0.4957$  & $0.9446$    & $1.0172$\\
    wet (sum)    & $0.4829$  & $0.5046$    & $1.0708$\\
    chewy (sum)  & $0.1667$  & $0.4179$    & $5.0000$\\
    \bottomrule
  \end{tabular}
\end{table}

Table \ref{tab:food_results_bouts_loso} presents LOSO and LOFTO results for the
bout-level algorithm. Comparing these results with chew-level results we can see
that the bout-level algorithm achieves almost similar results for {LOSO}:
slightly lower weighted accuracy for crispiness (but still quite high), and
almost the same for wetness. Recognition accuracy for chewiness is worse.

On the other hand, the bout-level algorithm seems to be able to better
generalize across different food types. Looking at the results of the {LOFTO}
experiment, the algorithm achieves only slightly lower weighted accuracy
(compared to the chew-level algorithm with majority voting) for crispiness and
improves significantly for wetness and chewiness.

\begin{table}
  \centering
  \caption{Results for bout-level recognition.}
  \label{tab:food_results_bouts_loso}
  \begin{tabular}{lcccc}
    \toprule
    & \textbf{Prior} & \textbf{Weighted accuracy} & \textbf{$w$}\\
    \midrule
    \emph{LOSO}\\
    crispy (avg) & $0.4967$   & $0.9541$    & $1.1030$\\
    crispy (sum) & $0.5084$   & $0.9534$    & $0.9669$\\
    wet (avg)    & $0.4869$   & $0.7865$    & $1.0880$\\
    wet (sum)    & $0.4958$   & $0.7900$    & $1.0169$\\
    chewy (avg)  & $0.1625$   & $0.5200$    & $6.0749$\\
    chewy (sum)  & $0.1597$   & $0.5238$    & $5.2632$\\
    \midrule
    \emph{LOFTO}\\
    crispy (sum) & $0.5084$   & $0.9288$    & $0.9669$\\
    wet (sum)    & $0.4958$   & $0.6422$    & $1.0169$\\
    chewy (sum)  & $0.1597$   & $0.4970$    & $5.2632$\\
    \bottomrule
  \end{tabular}
\end{table}

\section{Conclusions}
\label{sec:conclusions}

In this work we have proposed two algorithms for automatically and objectively
recognizing three attributes of food texture from audio signals captured by an
in-ear microphone. The algorithms combine feature extraction and binary SVMs and
operate both for single chews and entire chewing bouts. We have examined their
ability to generalize not only to new subjects (LOSO) but also to new food types
(LOFTO). With the use of these algorithms, the SPLENDID sensor was able to
recognize 3 important food-texture attributes affecting eating behavior. In
particular, crispiness was recognized with weighted accuracy of at least $0.9$
across all experiments. Results for recognizing wetness and chewiness are
promising but there is a large margin for improvement: introducing more suitable
features in the algorithms could possibly help improve the overall effectiveness
of the proposed algorithms. Including more food types in the training set could
also potentially improve recognition, however, certain food types with
attributes that are not easy to annotate (e.g. food types that are neither
completely dry nor wet) might not be suitable for a crisp-label classification
problem and thus require alternative methods. As a result, this type of digital
devices will make it possible to further study the objective exposure to
different food textures in relation to eating behavior and longer term outcomes,
such as weight change, in a real life setting.

\bibliographystyle{splncs04}
\bibliography{paper}

\end{document}